\pdfoutput=1

\documentclass[prd,nofootinbib,preprintnumbers,
  preprint,
  ]{revtex4-2}
\usepackage{amssymb,amsmath}
\usepackage{graphicx}
\usepackage[T1]{fontenc}
\usepackage{lmodern}
\usepackage{caption}
\usepackage{subcaption}
\usepackage{float}
\usepackage{pifont}
\usepackage{bm}
\usepackage{setspace}

\usepackage[colorlinks=true
  ,urlcolor=blue
  ,citecolor=blue
  ,linkcolor=blue
]{hyperref}

\usepackage{todonotes}

\begin{document}

\title{Light inflaton model in a metastable Universe}

\author{Fedor Bezrukov}
\email{Fedor.Bezrukov@manchester.ac.uk}
\author{Abigail Keats}
\email{Abigail.Keats@postgrad.manchester.ac.uk}

\affiliation{Department of Physics and Astronomy,
  University of Manchester, Manchester, M13 9PL, United Kingdom}

\date{September 2021}

\begin{abstract}
  We minimally extend the Standard Model (SM) with a Z$_2$ symmetric potential containing a single scalar field, serving as our inflaton with a quartic self-coupling. In the model we have symmetry breaking in both sectors, and with the addition of an inflaton-Higgs portal, the Universe is able to efficiently reheat via 2-2 inflaton-Higgs scattering. Assuming that the Universe with a positive cosmological constant should be metastable, only one particular symmetry breaking pattern in the vacuum is possible, without the need to finely-tune the Higgs quartic self-coupling. Inflatons with masses in the range  $O(10^{-3})\leq m_{\chi}\leq m_{h}$ and mixing angles that span $\theta_{m}^{2}=O(10^{-11}-10^{-2})$ evade all current cosmological, experimental and stability constraints required for a metastable electroweak (EW) vacuum. Upgraded particle physics experiments may be able to probe the parameter space with $\theta_{m}^{2}\geq O(10^{-4})$, where we would observe trilinear Higgs couplings suppressed by up to $2\%$ compared to the SM value. However to access the parameter space of very weakly-coupled inflaton, we rely on the proposals to build experiments that target the hidden sector.
\end{abstract}

\maketitle
\newpage

\section{Introduction}

The Standard Model (SM) is in excellent agreement with the experimental data at the present time. Moreover, the precision measurements of the Higgs boson mass, the top quark mass and the strong coupling constant indicate that the SM can be considered as a weakly coupled model up to the Planck scales. Intriguingly, the central measured values for the top and Higgs boson masses correspond to the situation of a metastable electroweak (EW) vacuum \cite{Bezrukov:2012sa,Elias_Mir__2012,Bednyakov_2015,PDG2020}. The SM, however, does not include mechanisms for inflation \cite{Starobinsky:1980te,Mukhanov:1981xt,Guth:1980zm,Linde:1981mu,Albrecht:1982wi}, and should therefore be extended.

The experimental indication of the metastability of the EW vacuum is especially interesting in view of the theoretical expectations of the inconsistency of eternal de Sitter expansion of the Universe, which would correspond to the present day dark energy domination \cite{dvali2014quantum,Dvali_2017}. As far as it is argued, that the eternal de Sitter expansion would lead to inconsistencies due to quantum breaking from graviton-graviton scattering, it is required that the expansion finishes in some, possibly rather far, future. Therefore, we require a mechanism by which the Universe can gracefully end before this critical time is reached. A Universe in a false-vacuum state with a sufficiently long lifetime (a metastable vacuum) will eventually decay into the true vacuum state, thus providing us with a solution to this problem. At the same time, the inflationary scenario should lead in some way to the creation of the Universe filled with this metastable state, to coincide with the current observations, in particular to our existence.

We extend the SM by an additional scalar inflaton field with a simple Z$_{2}$ symmetric potential, which allows the full model to be weakly coupled. The model can then provide us with both good inflationary behaviour while retaining metastability of the EW vacuum at present times. During inflation the Higgs direction is stabilized by its positive mass generated by the interaction with the inflaton field \cite{Kohri_2017,Lebedev_2013}, while at late time the contribution of the inflaton is negligible, and the EW vacuum becomes metastable. In addition, by demanding that we have successful preheating with a reheating temperature at least above the EW phase transition, which is required for the generation of baryon asymmetry from  leptogenesis, we arrive to a rather tightly constrained region of parameters for the inflaton field. In particular, this region will be largely explored by current and planned experiments on the high intensity frontier.

The article is organised as follows. In section~\ref{sec:Model} we outline the details of the model. Section~\ref{sec:EW} gives an overview of the metastable EW vacuum. Section~\ref{sec:Constraints} highlights the cosmological and observational constraints on the inflaton. In section~\ref{sec:Analytical} an analytical approximation of the allowed region for the experimentally observable parameters is given. Section~\ref{sec:Results} discusses the results.

\section{The Model}
\label{sec:Model}

A minimal extension of the SM that incorporates mechanisms for inflation and reheating can be achieved with the addition of a single scalar field serving as our inflaton, $X$, which couples to the SM Higgs doublet, $\Phi$. The Z$_{2}$ symmetric\footnote{A small cubic term, $\mu X^{3}$, may solve the domain wall problem without influencing the dynamics of inflation and reheating providing $\mu\lesssim\sqrt{\alpha^{2}/\lambda v}$ \cite{Anisimov_2009}.} model assumes parameters that will lead to symmetry breaking of the Higgs field, tuned to the SM expected value, and of the inflaton field, to evade a relic abundance of stable particles. The model includes a quartic scalar potential, which dominates the energy density of the Universe during slow-roll inflation; negative squared mass terms for the scalar and Higgs fields, giving rise to symmetry breaking in both sectors; and a scalar-Higgs portal coupling, required for reheating: 
\begin{align} \label{eq:V}
    V(X,\Phi) = \frac{\beta}{4}X^{4}-\frac{1}{2}\mu_{X}^{2}X^{2}-\mu_{\Phi}^{2}\Phi^{\dagger}\Phi+
    \lambda\Big(\Phi^{\dagger}\Phi+\frac{\alpha}{\lambda}X^{2}\Big)^{2}.
\end{align}
We consider both positive and negative $\alpha$ in our analysis.\footnote{Note, that in the previous works \cite{Shaposhnikov_2006,Anisimov_2009,Bezrukov_2010,Bezrukov_2020,Bezrukov_2020} the sign convention was the opposite to \eqref{eq:V}, i.e.\ $\alpha\equiv-\alpha_{\text{present work}}$.} As far as we are only interested in the situation where both fields have broken symmetry in order to correspond to observations, we can write the following formulas for $\mu_{\Phi}$ and $\mu_{X}$ for any non-zero vacuum expectation values (VEVs) of the fields, $v\equiv \sqrt{2}\langle\Phi\rangle$ and $v_{X}\equiv\langle X\rangle$:
\begin{align}
    \mu_{\Phi}^{2}&=v^{2}\lambda+2\alpha v_{X}^{2} ,
    \\
    \mu_{X}^{2}&=v_{X}^{2}\left(\beta+\frac{4\alpha^{2}}{\lambda}\right)+2\alpha v^{2} .
\end{align}
Going forward, we will work in the parameter space that trades $\mu_{\Phi}$ and $\mu_{X}$ in favour of $v$ and $v_{X}$, where $v$ is the well known SM Higgs VEV, equal to $246$ GeV. It may be convenient to introduce the ``vacuum angle'' as
\begin{align} \label{eq:thetaV}
    \tan\theta_{\text{V}} \equiv \frac{v}{v_{X}} .
\end{align}
Spontaneous symmetry breaking of the fields gives mass to excitations of the fields on top of the vacua in the mass basis $(\tilde{h},\tilde{\chi})$ that is rotated with respect to the gauge basis, $(h,\chi)\equiv(\sqrt{2}\Phi-v,X-\langle X\rangle)$:
\begin{align} \nonumber
    \tilde{h}    &= h\cos{\theta_{\text{m}}}-\chi\sin\theta_{\text{m}},
    \\ 
    \tilde{\chi} &= \chi\cos{\theta_{\text{m}}}+h\sin\theta_{\text{m}}.
\end{align}
As the angle $\theta_{\text{m}}$ is a complicated expression in terms of $v$ and $v_{X}$,  in equation (\ref{eq:thetam}) below we exceptionally use the mass of the physical inflaton state $\tilde{\chi}$, denoted by $m_{\chi}$, for which an approximate analytical formula (\ref{eq:mass}) is given later in paper:
\begin{align} \label{eq:thetam}
    \tan\theta_{\text{m}}=
    \begin{cases}
        +\left(\frac{2\alpha v_{X}}{v\lambda}\right)\frac{1}{1-\frac{m_{\chi}^{2}}{2v^{2}\lambda}}, & \alpha<0 \\
        -\left(\frac{2\alpha v_{X}}{v\lambda}\right)\frac{1}{1-\frac{m_{\chi}^{2}}{2v^{2}\lambda}}. & \alpha>0 
    \end{cases}
\end{align}

\section{Metastability of the electroweak vacuum}
\label{sec:EW}

The running of the Higgs quartic self-coupling, $\lambda$, to high energy scales determines the nature of the EW vaccum. If $\lambda$ is positive at all energy scales, the EW vacuum is stable as it is the only vacuum state. However, if $\lambda$ becomes negative at some energy scale (called the instability scale, $\mu_{s}$), there exists an additional vacuum state with lower-energy that the EW vacuum will ultimately decay into. In this case, the EW vacuum is said to be metastable if its lifetime exceeds the age of the Universe. Figures \ref{fig:LEn} and \ref{fig:Vu} below illustrate the renormalization group (RG) evolution of $\lambda$ and the SM Higgs potential respectively, for a stable EW vacuum (blue curves) and metastable EW vacuum (red curves). 
\begin{figure}
    \centering
    \begin{subfigure}{0.48\textwidth}
        \centering
        \includegraphics[width=\textwidth]{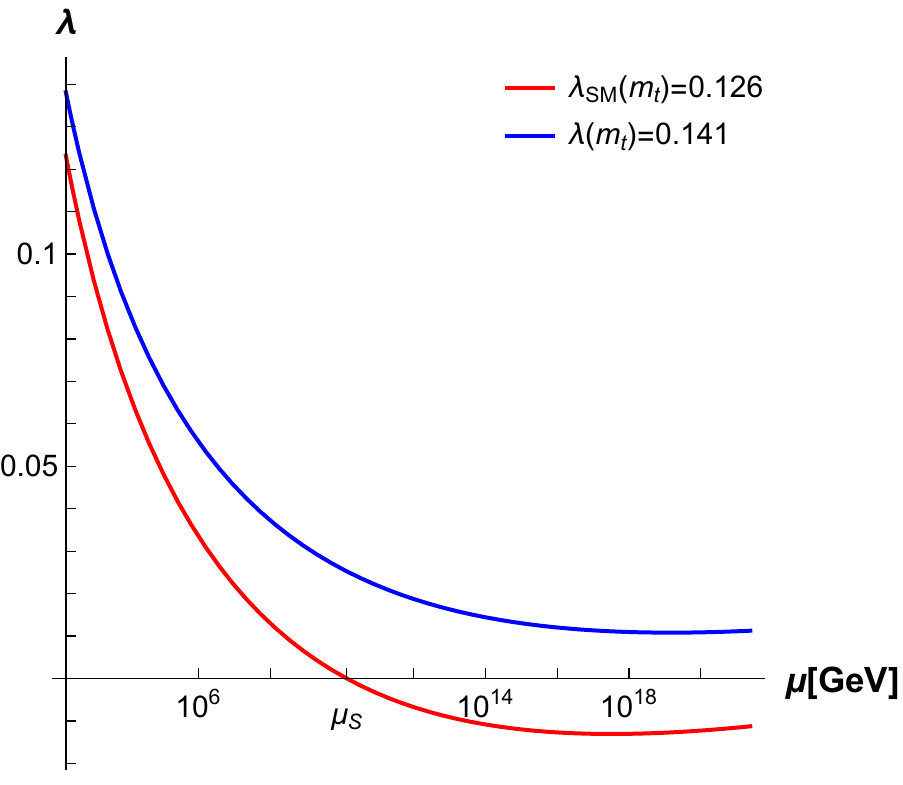}
        \caption{}
        \label{fig:LEn}
    \end{subfigure}
    \hfill
    \begin{subfigure}{0.48\textwidth}
        \centering
        \includegraphics[width=\textwidth]{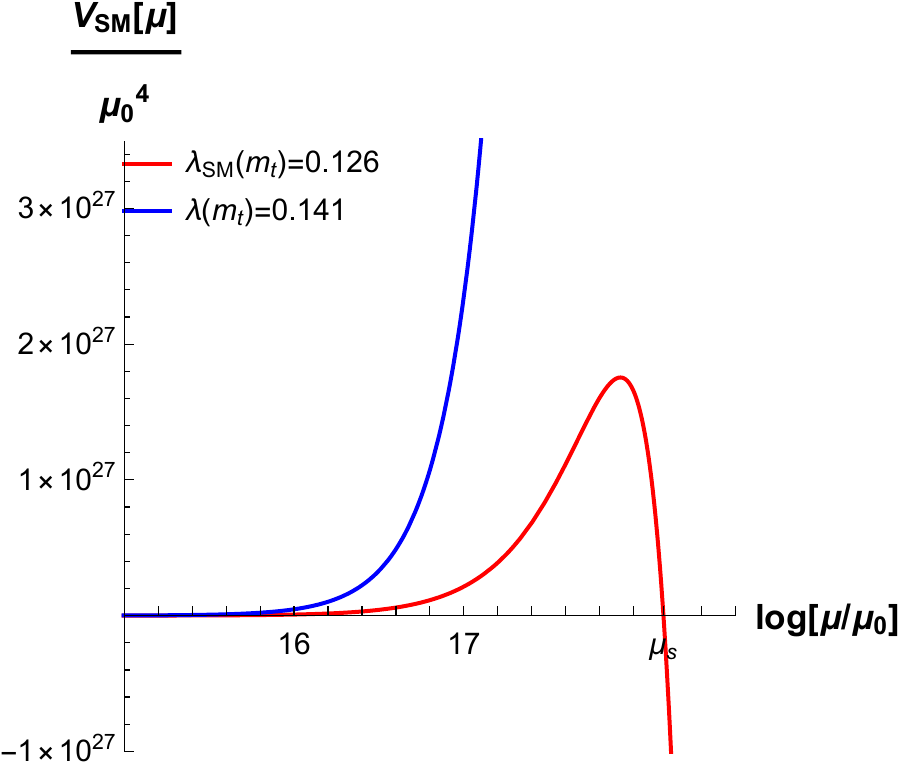}
        \caption{}
        \label{fig:Vu}
    \end{subfigure}
    \caption{(a) 3-loop RG evolution of the Higgs self-coupling, $\lambda$, with energy scale, $\mu$, evaluated at the SM value $\lambda_{\text{SM}}(m_{t})=0.126$ in red and at $\lambda(m_{t})=0.141$ in blue. The instabilty scale, $\mu_{S}=O(10^{10})$ GeV, is the energy scale where $\lambda_{\text{SM}}$ becomes negative. (b) The Higgs quartic potential, $V_{\text{SM}}[\mu]=\frac{\lambda[\mu]}{4}\mu^{4}$, normalised at the reference scale $\mu_{0}=m_{t}$.}
    \label{fig:EW}
\end{figure}
The running of $\lambda$ is especially sensitive to experimental inputs of the top quark mass, $m_{t}$, and the strong coupling constant, $\alpha_{s}$, due to the large negative contribution of the top Yukawa coupling, $y_{t}$, to $\lambda$'s beta function. At 1-loop the beta functions of $\lambda$ and $y_{t}$ are\footnote{Modification from the inflaton, $X$, is negligible; $\delta\beta_{\lambda}\sim\alpha^{2}\lesssim 10^{-10}$.} \cite{Chetyrkin_2012}:
\begin{align}
\label{eq:betaL}
    \beta_{\lambda}&=\frac{1}{16\pi^{2}}\Big(12\lambda^{2}+6y_{t}^{2}\lambda-3y_{t}^{4}\Big),
    \\ 
    \label{eq:betay}
      \beta_{y_{t}}&=\frac{1}{16\pi^{2}}\Big(\frac{9}{4}y_{t}^{3}-4g_{s}^{2}y_{t}\Big). 
\end{align}
Smaller $m_{t}$ and larger $\alpha_{s}$ increase $\lambda$ and therefore stabilise the EW vacuum. We use the following experimental values of the top quark's pole mass and $\alpha_{s}(m_{Z})$, quoted with one standard deviation from their central value \cite{PDG2020}:
\begin{align}
    \label{eq:mt}
    m_{t}             &= 172.76\pm0.30 \text{ GeV},
    \\ 
    \label{eq:as}
    \alpha_{s}(m_{Z}) &= 0.1179\pm0.0010.
\end{align}
The measurement of the Higgs boson mass in the SM fixes its' self-coupling at the EW scale $[\lambda_{\text{SM}}(m_{h})=0.1291\pm0.0003$  \cite{PDG2020}$]$. The central values thus correspond to the mateastable EW vacuum \cite{Bednyakov_2015}. In our model, however, the value of $\lambda$ is not uniquely fixed by the Higgs boson mass. Therefore, to assure the Universe is in the metastable state at present, as outlined in the introduction, we just demand $\lambda$ to be within the stability bound, $\lambda<\lambda_{\text{stab}}$
\cite{Bednyakov_2015}:
\begin{align}
    \label{eq:stab}
    \lambda_{\text{stab}}(m_{t})=0.134 .
\end{align}
Additionally $\lambda$ is constrained by the instability bound, $\lambda>\lambda_{\text{instab}}$ \cite{Espinosa_2008,Isidori_2001}:
\begin{align}
    \label{eq:instab}
    \lambda_{\text{instab}}(m_{t})=0.1 ,
\end{align}
by the requirement that our Universe exists at all; otherwise, the Universe would decay faster than its lifetime. Both bounds (\ref{eq:stab}) and (\ref{eq:instab}) are evaluated at the central values of $m_{t}$ (\ref{eq:mt}) and $\alpha_{s}$ (\ref{eq:as}), with the renormalization scale equal to the top quark mass.

However, the bounds (\ref{eq:stab}) and (\ref{eq:instab}) alone do not ensure the survival of a mestable Universe throughout the course of its history. In particular, periods of inflation and reheating could easily destabilise the metastable vacuum, and so additional model-dependent constraints are necessary. In addition to metastability constraints, those from cosmology and ensuring consistency of the model with current experimental measurements of the Higgs boson are the subjects of the following section.

\section{Constraints}
\label{sec:Constraints}

\subsection{Cosmological constraints}

\subsubsection{Inflationary constraints}

During the slow-roll inflationary epoch, we decompose the fields into a homogeneous classical component  and a quantum component,
\begin{align}
    X(x,t)&=X_{b}(t)+\chi(x,t),
    \\ \nonumber
    \Phi(x,t)&=\Phi_{b}(t)+\frac{h(x,t)}{\sqrt{2}}.
\end{align} 
The classically evolving background fields converge towards the inflationary attractor solution, which is found by evaluating the gradient of the potential, $\frac{\partial V}{\partial \theta_{\text{inf}}}=0$, using the following field transformations,
\begin{align}\nonumber
    X_{b}&\rightarrow R\cos\theta_{\text{inf}},
    \\ 
    \sqrt{2}\Phi_{b}&\rightarrow R\sin\theta_{\text{inf}},
\end{align}
where $R$ corresponds to the coordinate along the inflationary direction. The direction of the inflationary attractor solution in the field space is then given by \cite{Bezrukov_2020},
\begin{align}
\label{eq:IA}
    \tan\theta_{\text{inf}}\equiv\frac{\sqrt{2}\Phi_{b}}{X_{b}}\sim
        \sqrt{\frac{\beta-2\alpha}{\lambda}} ,
\end{align}
requiring that $|\alpha|,\beta\ll\lambda$. To establish if slow roll inflation is compatible with a metastable Universe, we evaluate $\Phi_{b}$ for $\alpha\lessgtr0 \text{ and } 2|\alpha|\lessgtr\beta)$ in Table \ref{Tab:thetainf}.
\begin{table}[H]
    \begin{center}
    \begin{tabular}{ |c||c|c| } 
    \hline
        & \bm{$\alpha>0$}& \bm{$\alpha<0$} 
    \\ 
    \hline
    \hline
        \bm{$2|\alpha|>\beta$}& $\sqrt{2}\Phi_{b}=0$ &$\sqrt{2}\Phi_{b}\sim\sqrt{\frac{2|\alpha|}{\lambda[\mu]}}X_{b}$  
    \\ 
    \hline
        \bm{$2|\alpha|<\beta$} & $\sqrt{2}\Phi_{b}\sim\sqrt{\frac{\beta}{\lambda[\mu]}}X_{b}$&  $\sqrt{2}\Phi_{b}\sim\sqrt{\frac{\beta}{\lambda[\mu]}}X_{b}$
    \\
    \hline
    \end{tabular}
    \caption{\label{Tab:thetainf}$\Phi_{b}$ is evaluated from equation (\ref{eq:IA}) for $\alpha\lessgtr0 \text{ and } |\alpha|\lessgtr\beta$.}
    \end{center}
\end{table}
In regions of parameter space where $\theta_{\text{inf}}\neq 0$, we require the initial condition that $\Phi_{b}(0)$ starts from a point in field space where $\lambda[\Phi_{b}]>0$. $\Phi_{b}(t)$ then evolves to larger field values as it converges towards the inflationary attractor solution, which simultaneously increases as $\lambda$ runs to higher energy scales. For most of the metastability parameter space, $\Phi_{b}$ evolves to field values of $O(10^{16}-10^{18})\text{ GeV}$, which greatly exceed the instability scale, thus $\lambda[\Phi_{b}]$ becomes negative and the inflationary attractor breaks-down. Only $\lambda$ within $\sim1\%$ of the stability bound (\ref{eq:stab}) gives $\Phi_{b}<\mu_{s}$ during inflation. Additionally for $\alpha<0$, the negative cross-term is large enough to dominate the Higgs potential up to the scale of $O(M_{\text{P}})$ if
\begin{equation}
    \label{eq:negcross}
    \alpha\gtrsim O\Bigg(10^{-10} \times \bigg(\frac{\mu_{s}[\text{GeV}]}{10^{16}}\bigg)^{2}\Bigg) .
\end{equation}
In this region, the potential is unstable at the origin and so the Higgs field can only evolve towards the vacuum state beyond the scale of $M_{\text{P}}$. We conclude that a non-zero inflationary attractor is only compatible with slow-roll inflation in a metastable Universe for finely-tuned $\lambda\sim\lambda_{\text{stab}}$ in the parameter space $\alpha>0 \ \& \ 2|\alpha|<\beta$, and $\alpha<0$ if $|\alpha|$ is smaller than the destabilising region defined in (\ref{eq:negcross}). The focus of our analysis going forward will therefore be on constraining the region $\alpha>0 \ \& \ 2|\alpha|>\beta$, where for $\Phi_{b}=0$, we evade the need to finely tune $\lambda$.

For the model to be consistent with the Cosmic Microwave Background (CMB) measurement of the primordial scalar density perturbation amplitude \cite{Lyth_1999} and within the tensor-to-scalar ratio limit $(r<0.13)$ \cite{2020}, we require a non-minimal coupling of the scalar field to gravity, $\xi X^{2}R/2$ \cite{Kaiser_1995,Komatsu_1999,Bezrukov_2013}. Through a conformal transformation of the metric, the following expression for $r$ is obtained \cite{Bezrukov_2013,Anisimov_2009}:
\begin{align}
    r=\frac{16\left(1+6\xi\right)}{\left(N+1\right)\left(1+8\left(N+1\right)\xi\right)},
\end{align}
where for $\xi\geq O(10^{-2})$, $r$ is sufficiently suppressed when evaluated at the relevant number $e$-foldings prior to the end of inflation for our model, $N\sim 60$. If we assume no new scales below the Plank scale $(\xi<1)$, we obtain the following range for the scalar field's quartic self-coupling \cite{Bezrukov_2013,Anisimov_2009}:
\begin{equation} \label{eq:b}
    O(10^{-12}) \leq \beta \leq O(10^{-9}) ,
\end{equation}
where the lower bound corresponds to the excessive production of tensor modes and the upper bound would require $\xi>1$ and introduces additional scales below Planck.

Additionally, to ensure quantum corrections to the inflaton’s quartic self-coupling are sufficiently small and the inflationary analysis above holds, we require $\alpha^{2}<0.1\beta$ \cite{Shaposhnikov_2006}, leading to
\begin{equation} \label{eq:ba1}
    \alpha\leq (0.1\beta)^{\frac{1}{2}}.
\end{equation}

\subsubsection{Constraints from preheating}

For efficient sphaleron conversion of lepton to baryon asymmetry, we require the reheating temperature to exceed the electroweak symmetry breaking scale, $T_{\text{EW}} = 160$ GeV \cite{D_Onofrio_2016}. The minimum reheating temperature requirement translates to a lower bound on $\alpha$, which has been established previously for the model with $\mu_{\Phi}^{2}=0$. The analyses are carried out separately for $m_{\chi}<2m_{h}$ \cite{Shaposhnikov_2006,Anisimov_2009,Bezrukov_2010} and  $m_{\chi}>2m_{h}$ \cite{Bezrukov_2020}, as they reheat via different mechanisms: $\chi\chi\rightarrow hh$ and $\chi\rightarrow hh$ respectively. However inflation requires our model to have $\alpha>0 \ \& \  2|\alpha|>\beta$ (for $\lambda$ that is not finely-tuned), which limits us to light inflaton, $m_{\chi}\lesssim m_{h}$. 

Following inflation is a preheating period, during which parametric resonance excites inflaton particles to occupy a highly infra-red distribution but does not efficiently transfer energy into the SM (due to $\alpha\ll\lambda$). The inflaton distribution slowly evolves self-similarly towards thermal equilibrium \cite{Micha_2004,PhysRevLett.90.121301} until perturbative reheating proceeds once the mean-free path is comparable to the Hubble expansion rate, $n_{\chi}\sigma_{\chi\chi\rightarrow hh}\sim H$ \cite{Anisimov_2009}. We assume at this moment the inflatons have not yet thermalised and so their average momentum is suppressed\footnote{The exponent given in equation (\ref{eq:pavgT}) assumes that 4-particle scatterings drive the evolution of the distribution \cite{Micha_2004,PhysRevLett.90.121301}.} with respect to the thermal estimate, $p_{\text{avg}}/T\sim1$:
\begin{align}
\label{eq:pavgT}
    \frac{ p_{\text{avg}}}{T}&=\beta^{\frac{2}{7}}\bigg(\frac{M_{\text{P}}}{T}\bigg)^{\frac{1}{7}},
\end{align}
thereby enhancing the mean-free path and increasing the reheating  temperature with respect to the thermal estimate \cite{Anisimov_2009}, 
\begin{align}
\label{eq:thermalT}
T_{\text{reh,T}}\approx\frac{\zeta(3)\alpha^{2}}{\pi^{4}}\sqrt{\frac{90}{g_{\text{SM}}}}M_{\text{P}} ,
\end{align}
 by a factor of $(T/p_{\text{avg}})^{3}$. The thermal estimate of the lower bound on $\alpha$ is
\begin{align}
    \alpha>\alpha_{\text{min,T}}\approx\left(\frac{\pi^{4}}{\zeta(3)}\sqrt{\frac{g_{\text{SM}}}{90}}\frac{T_{\text{EW}}}{M_{\text{P}}}\right)^{\frac{1}{2}}\sim 7.5\times 10^{-8}.
\end{align}
In our analysis we will use the following more realistic estimate that assumes a non-thermal inflaton distribution,
\begin{equation} \label{eq:ba}
    \alpha > \alpha_{\text{min}} \approx\left(\frac{M_{\text{P}}}{T_{\text{EW}}}\right)^{\frac{3}{14}}\left(\frac{\pi^{4}}{\zeta(3)}\sqrt{\frac{g_{\text{SM}}}{90}}\frac{T_{\text{EW}}}{M_{\text{P}}}\right)^{\frac{1}{2}}\beta^{\frac{3}{7}}  \sim (1.6\times 10^{-9}) \left(\frac{\beta}{10^{-12}}\right)^{\frac{3}{7}},
\end{equation}
with $g_{\text{SM}}\sim 100$. 
Together with \eqref{eq:ba1} this fully closes the $\beta-\alpha$ parameter space, as summarized in Table \ref{tab:ba}.
\begin{table}[H]
    \centering
    \begin{tabular}{ |c|c| } 
    \hline
        \bm{$\beta$}  & \bm{$\alpha$}   \\ 
    \hline
    \hline
        $10^{-9}$ & $(3.1\times10^{-8})-(1.0\times10^{-5})$ \\ 
    \hline
        $10^{-10}$ &  $(1.1 \times 10^{-8})-(3.2\times10^{-6})$ \\ 
    \hline
        $10^{-11}$ & $(4.2\times10^{-9})-(1.0\times10^{-6})$ \\ 
    \hline
        $10^{-12}$ & $(1.6\times10^{-9})-(3.2\times10^{-7})$  \\ 
    \hline
    \end{tabular}
    \caption{\label{tab:ba}$\beta$-$\alpha$ parameter space allowing successful inflation and preheating.}
\end{table}

\subsubsection{Big Bang Nucleosynthesis constraints}

The measurement of primordial elemental abundances tightly constrains the number of additional relativistic degrees of freedom (approximately less than half a neutrino specie) to the SM at this epoch \cite{Gorbunov_2017}.  Light inflaton should therefore preferably decay prior to big bang nucleosynthesis (BBN), $\Gamma_{\chi}^{-1}\lesssim 1$ s, which constrains the mixing angle from below:
\begin{align}
    \Gamma_{\chi}&=\sin^{2}\theta_{\text{m}}\Gamma_{h}(m_{\chi}),
    \\ 
    \label{eq:BBNbound}
    \theta_{\text{m}}&\gtrsim\sqrt{\frac{\Gamma_{h}^{-1}(m_{\chi})}{1\text{ s}}}.
\end{align}
Here $\Gamma_{h}(m_{\chi})$ is the SM Higgs boson decay width evaluated if its mass is equal to $m_{\chi}$ \cite{Bezrukov_2013}. The lower bound on $\theta_{\text{m}}$ is evaluated for $m_{\chi}\leq1$ GeV in Table \ref{Tab:theta}.
\begin{table}[H]
    \begin{center}
    \begin{tabular}{ |c|c| } 
    \hline
        \bm{$m_{\chi}$}\textbf{[GeV]} & \bm{$(\theta_{\text{\textbf{m}}}^{\text{\textbf{BBN}}})^{2}$} 
    \\ 
    \hline
    \hline
        $10^{-4}$ & $0.44$ \\
    \hline
        $10^{-3}$ & $3.3\times 10^{-4}$ 
    \\ 
    \hline
        $10^{-2}$ & $3.9\times 10^{-10}$
    \\
    \hline
        $10^{-1}$ & $3.8\times 10^{-11}$  
    \\ 
    \hline
    \end{tabular}
    \caption{\label{Tab:theta}Minimum $\theta_{\text{m}}$ for a given inflaton mass, to ensure decay prior to BBN.}
    \end{center}
\end{table}

\subsubsection{$\langle X\rangle=0$ is excluded by overclosure of the Universe}

The relic inflaton with $\langle X\rangle=0$, which cannot deplete via decay into the SM, freezes out with a large abundance. This could be a possible source of dark matter (DM) with the inflaton playing the role of a weakly interacting massive particle (WIMP), but this situation is fully excluded, as it leads to overclosure of the Universe. 

Post reheating the inflaton is in thermal equilibrium with the SM bath; however, as the Universe expands and cools, the inflaton decouples and freezes out once its annihilation rate drops below the Hubble expansion rate. Inflaton that are non-relativistic at the time of freeze-out, $T_{f}< m_{\chi}$, have a relic abundance today of \cite{Gorbunov_2017}
\begin{align}
\label{OmegaX}
    \Omega_{\chi}\sim O(10^{-10})\Bigg(\frac{\text{GeV}^{-2}}{\langle\sigma_{\text{ann}}v\rangle}\Bigg).
\end{align}
The dominant contribution to the light inflaton cross-section, $m_{\chi}\ll m_{h}$, is the Higgs-mediated $s$-channel annihilation into SM particles,
\begin{align}
\label{css}
    \langle\sigma_{\text{ann}} v\rangle_{\chi\chi\rightarrow\text{SM,SM}} \sim \frac{4\alpha^{2} v^{2}}{m_{\chi}m_{h}^{4}}\Gamma_{h}(2m_{\chi})
    \lesssim O(10^{-20}) \text{ GeV}^{-2} , 
\end{align}
for $ m_{\chi}\leq1 \text{ GeV}$ and $\alpha\leq 10^{-5}$. For heavy inflaton, $m_{\chi}\gg m_{h}$, there is an additional contribution from the direct 2-2 inflaton-Higgs vertex,
\begin{align}
\label{csl}
    \langle\sigma_{\text{ann}} v\rangle_{\chi\chi\rightarrow\text{SM,SM}}+\langle\sigma_{\text{ann}} v\rangle_{\chi\chi\rightarrow\text{hh}}
    &\sim \frac{\alpha^{2} v^{2}}{4m_{\chi}^{5}}\Gamma_{h}(2m_{\chi})+ \frac{\alpha^{2}}{16\pi m_{\chi}^{2}} ,
    \\ \nonumber
    &\lesssim {O(10^{-17})\text{ GeV}^{-2}} ,
\end{align}
for $m_{\chi}\leq10^{3}$ GeV and $\alpha\leq 10^{-5}$. In both cases, (\ref{css}) and (\ref{csl}), the inflatons overclose the Universe, $\Omega_{\chi}\gg\Omega_{\text{DM}}$. Very light inflatons are relativistic at the time of freeze-out, $T_{f}\gg m_{\chi}$, and have a relic abundance today of \cite{Zyla:2020zbs,Gorbunov_2017}
\begin{align}
    \Omega_{\chi}\sim\Omega_{\text{DM}}\Bigg(\frac{m_{\chi}}{16 \text{ eV}}\Bigg) ;
\end{align}
we assume $1\text{ MeV}\lesssim T_{f}\lesssim 100 \text{ MeV}$, corresponding to $g_{*}(T_{f})=10.75$ effective degrees of freedom. BBN constraints disfavour $m_{\chi}\ll T_{\text{BBN}}\sim 1$ MeV, so we conclude that very light inflatons also overclose the Universe.

\subsection{Particle physics constraints}

\subsubsection{Higgs measurement constraints}

Our model must also be consistent with the current experimental measurement of the Higgs boson's signal strength and the invisible decay width bound. The signal strength, $\mu$, is defined as the product of the Higgs boson production cross-section and its branching ratio ($\sigma \cdot BR$) observed, normalised to that of the SM. The result from combined ATLAS and CMS Run 1 data is \cite{PhysRevD.98.030001}
\begin{align}
    \mu &\equiv \frac{(\sigma\cdot BR)_{\text{obs}}}{(\sigma\cdot BR)_{\text{SM}}}
        = 1.09\pm0.11 ,
\end{align}
from which we obtain the following $1\sigma$ upper bound on the mixing angle $\theta_{\text{m}}$,
\begin{align} \label{eq:theta1}
    |\theta_{\text{m}}|<|\theta_{1}|=0.14 .
\end{align}
In the parameter space $m_{\chi}<m_{h}/2$, there is an additional bound from the Higgs boson's invisible branching ratio, $BR_{\text{inv}}$, 
\begin{align}
 \label{eq:gammabound}
    \Gamma_{h\rightarrow\chi\chi}&\leq\Bigg(\frac{1}{1-BR_{\text{inv}}}-\cos^{2}\theta_{\text{m}}\Bigg)\Gamma_{SM} ;
\end{align}
where  $\Gamma_{\text{SM}}=4.1$ MeV is the theoretical SM Higgs boson width and
\begin{align}
\label{eq:gammahxx}
     &\Gamma_{h\rightarrow \chi\chi}\simeq\frac{\alpha^{2}v^{2}}{2\pi m_{h}}\bigg(1+6\frac{(\lambda_{\text{SM}}-\lambda)^{\frac{1}{2}}}{\lambda}\bigg)^{2} ,
\end{align}
evaluated in the limit $\mu_{\Phi}^{2}\gg v^{2}\lambda$ and $m_{\chi}\ll m_{h}$.
The $95\%$ C.L. limit on the invisible branching ratio from combined ATLAS Run 1 and Run 2 data is $BR_{\text{inv}}=0.26$ \cite{Aaboud_2019}, which imposes a weaker bound on our parameter space than (\ref{eq:theta1}). Although this bound will not contribute to our results, this analysis may be useful for future reference. 

\subsubsection{Direct detection constraints}

Furthermore, it is possible to directly constrain the inflaton particle created in high intensity experiments, which either escapes the detector (invisible mode) or decays later into a pair of observable particles.  The constraints on a light scalar boson with the Higgs mixing angle, $\theta_{\text{m}}$, and mass, $m_{\chi}$, can readily be used \cite{LHCb:2016awg,LHCb:2015nkv,E949:2008btt,Winkler_2019,Turner:1987by,PhysRevD.36.2201,PhysRevD.39.1020,PhysRevD.82.113008,Beacham_2019,Filimonova_2020,Gorbunov_2021} and are shown in Figures \ref{fig:mTb9} and \ref{fig:mTb13}.

\subsection{Stability Constraints}

\subsubsection{Inflation}

To prevent the Universe from being trapped in the true vaccum state, we require that the Higgs field value remains below the potential's instability scale, $\mu_{V}$, defined as the energy scale where the Higgs potential is zero. Although the inflationary attractor rapidly converges the Higgs background field to zero, we also need to ensure quantum fluctuations of the Higgs field, $h$, during inflation do not destabilise the EW vacuum. 

The effective mass of $h$ is dominated by the large inflaton field variance during slow-roll inflation.  Evaluated at 60 $e$-foldings prior to the end of inflation, $X_{60}=O(10M_{\text{P}})$, when the largest vacuum fluctuations are produced, 
\begin{align}
    m_{\text{eff}}=\sqrt{3\lambda\langle\Phi_{60}^{2}\rangle+2\alpha\langle X_{60}^{2}\rangle}\sim O(\sqrt{\alpha}\times 10^{19})\text{ GeV} ;
\end{align}
for simplicity we ignore the contribution from the backreaction of the Higgs field here. The large $m_{\text{eff}}$ stabilisies the EW vacuum by pushing the potential's instability scale to higher energy scales with respect to the SM\footnote{For the SM Higgs $\mu_{V}=\mu_{s}$, however this is not the case for our model due to the addition of the inflaton-Higgs coupling.} \cite{Kohri_2017,Lebedev_2013}
\begin{align}
\label{eq:uV}
    \mu_{V}\sim\frac{m_{\text{eff}}}{\sqrt{|\lambda|}}\gtrsim O(10^{15}-10^{17})\text{ GeV} ,
\end{align}
where $|\lambda|\leq O(0.01)$. As $m_{\text{eff}}$ is greater than the Hubble expansion rate\footnote{$U(X_{60})$ is the conformally transformed potential, which is a decreasing function of $\beta$, evaluated in the range $\beta=O(10^{-12}-10^{-9})$ \cite{Bezrukov_2013}.} \cite{Bezrukov_2013}, 
\begin{align}
    H(X_{60})=\sqrt{\frac{U(X_{60})}{3 M_{\text{P}}^{2}}}&=O(10^{13}-10^{14})\text{ GeV} ,
\end{align}
the amplitude of inflationary enhanced quantum fluctuations of the Higgs field is \cite{Lebedev_2013,Kohri_2017}
\begin{align}
\label{eq:qf}
   h=O\bigg(\frac{H(X_{60})}{10}\bigg)\sim O(10^{13}-10^{12})\text{ GeV} .
\end{align}
So we conclude that the EW vacuum is not destabilised during inflation, as we have shown $\mu_{V}>\Phi$ across the entire parameter space where $\alpha>0 \ \& \ 2|\alpha|>\beta$.

\subsubsection{Preheating and reheating}

Post-slow-roll inflation, the oscillatory zero-mode inflaton transfer energy into excitations of the fields through parametric resonance, which could destabilize the EW vacuum if the Higgs fluctuations exceed the instability scale. However, this is evaded as the Higgs scattering rate greatly exceeds the production rate, $\lambda\gg\alpha$, thereby promptly halting the transfer of energy into excitations of the Higgs field  \cite{Bezrukov_2020}. 

In the perturbative reheating regime, we require temperatures to exceed the EW symmetry breaking scale. Assuming the inflatons are non-thermal, (\ref{eq:pavgT}) and  (\ref{eq:thermalT}) are used to estimate the reheating temperature,
\begin{align}
    T_{\text{reh}}&\sim \bigg(\frac{T}{p_{\text{avg}}}\bigg)^{3}T_{\text{reh,T}}\sim\Big(\frac{\alpha}{\alpha_{\text{min}}}\Big)^{\frac{7}{2}}T_{\text{EW}} , 
    \\ 
    \label{eq:Trange}
    &=O(T_{\text{EW}}-10^{10})\text{ GeV} ,
\end{align}
for the range of $\alpha$ given in Table \ref{tab:ba}. Relatively low reheating temperatures stabilize the Higgs potential due to the addition of the thermal Higgs mass \cite{Salvio_2016},
\begin{align}
    V_{eff}(T,h)     &\approx V_{eff}(0,h)+\frac{m_{\text{T}}^{2}}{2}h^{2},
    \\ \nonumber
    m_{\text{T}}^{2} &\sim \Big(\frac{\lambda}{2}+\frac{g_{t}^{2}}{4}\Big) T^{2} ,
\end{align}
as positive thermal corrections increase the instability scale. However very high temperatures induce local nucleation of bubbles that probe the instability region, either via quantum tunnelling from an excited state or classical excitation over the potential barrier. The bubbles then expand rapidly, close to the speed of light, and destroy everything in their way \cite{Espinosa_2008,Elias_Mir__2012,Rose_2016}. However, cutting the parameter space at the $1\sigma$ Higgs' signal strength bound (\ref{eq:theta1}) restricts $\lambda$ to the range $0.123<\lambda(m_{t})\leq\lambda_{\text{SM}}$, and for
\begin{align}
    \label{eq:reh}
    \lambda(m_{t})>0.120,
\end{align}
the EW vacuum is sufficiently stable up to $T_{\text{reh}}=O(M_{\text{P}})$ \cite{Espinosa_1995,Espinosa_2008,Bednyakov_2015}.  We therefore assume the Universe can safely reheat; however, if a less stringent Higgs signal strength bound was to be considered, cuts from reheating may be necessary to ensure thermal fluctuations do not destabilise the EW vacuum.

\section{Analytical approximations}
\label{sec:Analytical}

The analytical approximations for the inflaton and Higgs boson masses for $\alpha>0 \ \& \ 2|\alpha|>\beta$ are:
\begin{align}
\label{eq:mass}
    m_{\chi}^{2}&\approx2\beta v_{X}^{2}-\frac{\beta^{2}v_{X}^{4}}{2\lambda v^{2}}+O(\alpha,\beta,\alpha^{2}) ,
    \\ \nonumber
    m_{h}^{2}&\approx 2\lambda v^{2}+\frac{\beta^{2}v_{X}^{4}}{2\lambda v^{2}}+O(\alpha,\beta,\alpha^{2})  =2\lambda_{\text{SM}}v^{2} ,
\end{align}
where the SM Higgs quartic coupling
is $\lambda_{\text{SM}}(m_{h})=0.129$. By fixing the Higgs mass, we obtain the $\lambda-v_{X}$ parameter space, for a given value of $\alpha$ and $\beta$,
\begin{align}
    \lambda_{\pm}=\frac{\lambda_{\text{SM}}}{2}\Bigg(1\pm\sqrt{1-\frac{4\beta^{2}v_{X}^{4}}{m_{h}^{4}}}\Bigg) .
\end{align}
In the limit $v_{X}\ll\frac{m_{h}}{\sqrt{2\beta}}$, the two solutions are
\begin{align}
\label{eq:lambda}
\begin{cases}
   \lambda_{+}\sim\lambda_{\text{SM}}\left(1-\left(\frac{\beta v_{X}^{2}}{m_{h}^{2}}\right)^{2}\right) , & \\
    \lambda_{-}\sim\lambda_{\text{SM}}\left(\frac{\beta v_{X}^{2}}{m_{h}^{2}}\right)^{2},            & 
\end{cases}
\end{align}
however $\lambda_{+}$ is the only solution that lies within the metastability bounds (\ref{eq:stab}) and (\ref{eq:instab}). The maximum value of $v_{X}$ is when $\lambda=\frac{\lambda_{\text{SM}}}{2}$,
\begin{align}
\label{eq:maxX}
    v_{X,\text{max}}& \approx\frac{m_{h}}{\sqrt{2\beta(1+Y)}}
\end{align}
where $Y=\frac{16\alpha^{2}}{\beta\lambda_{\text{SM}}}$. For $Y<1$, $v_{X,\text{max}}$ is a decreasing function of $\beta$, and for $Y\gtrsim1$, $v_{X,\text{max}}$ is a decreasing function of $\alpha$. Using approximations (\ref{eq:mass}) and (\ref{eq:maxX}), we can then see that the maximum inflaton mass approaches the Higgs mass for the smallest $Y$ and is suppressed for a larger $Y$,
\begin{align}
\label{eq:mxmax}
     m_{\chi,\text{max}}&\approx\frac{m_{h}}{\sqrt{1+Y}}\sim
\begin{cases}
    m_{h} , & \text{if } Y\ll 1\\
    \sqrt{\frac{\beta}{8\alpha^{2}}}\lambda_{\text{SM}}v ,              & \text{if }Y>1
\end{cases}.
\end{align}

\section{Results}
\label{sec:Results}

\begin{figure}
       \centering
     \begin{subfigure}{0.9\textwidth}
     \centering
     \includegraphics[width=\textwidth]{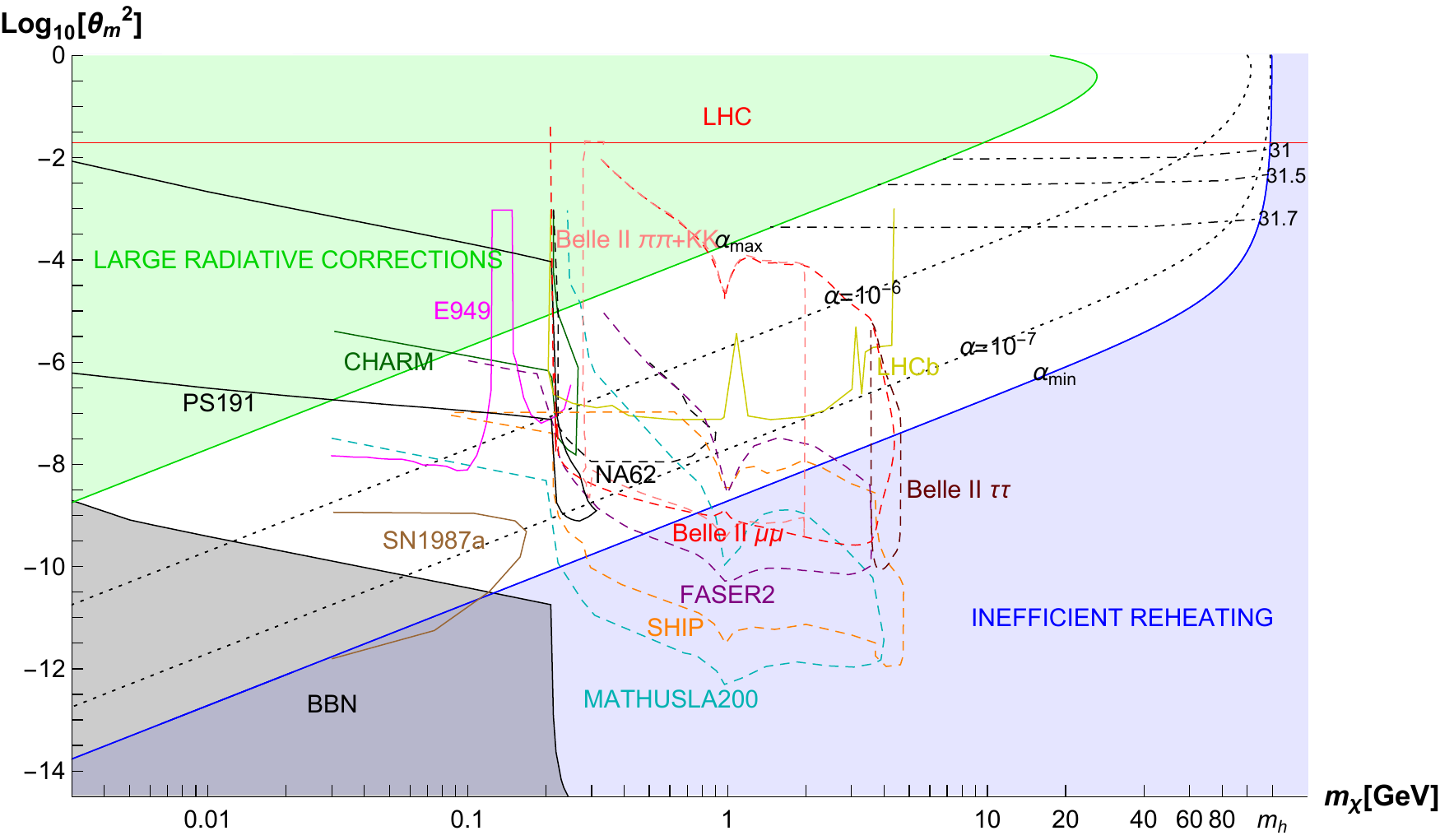}
         \caption{$\beta=10^{-9}$}
         \label{fig:mTb9}
     \end{subfigure}
     \hfill
     \begin{subfigure}{0.9\textwidth}
     \centering
     \includegraphics[width=\textwidth]{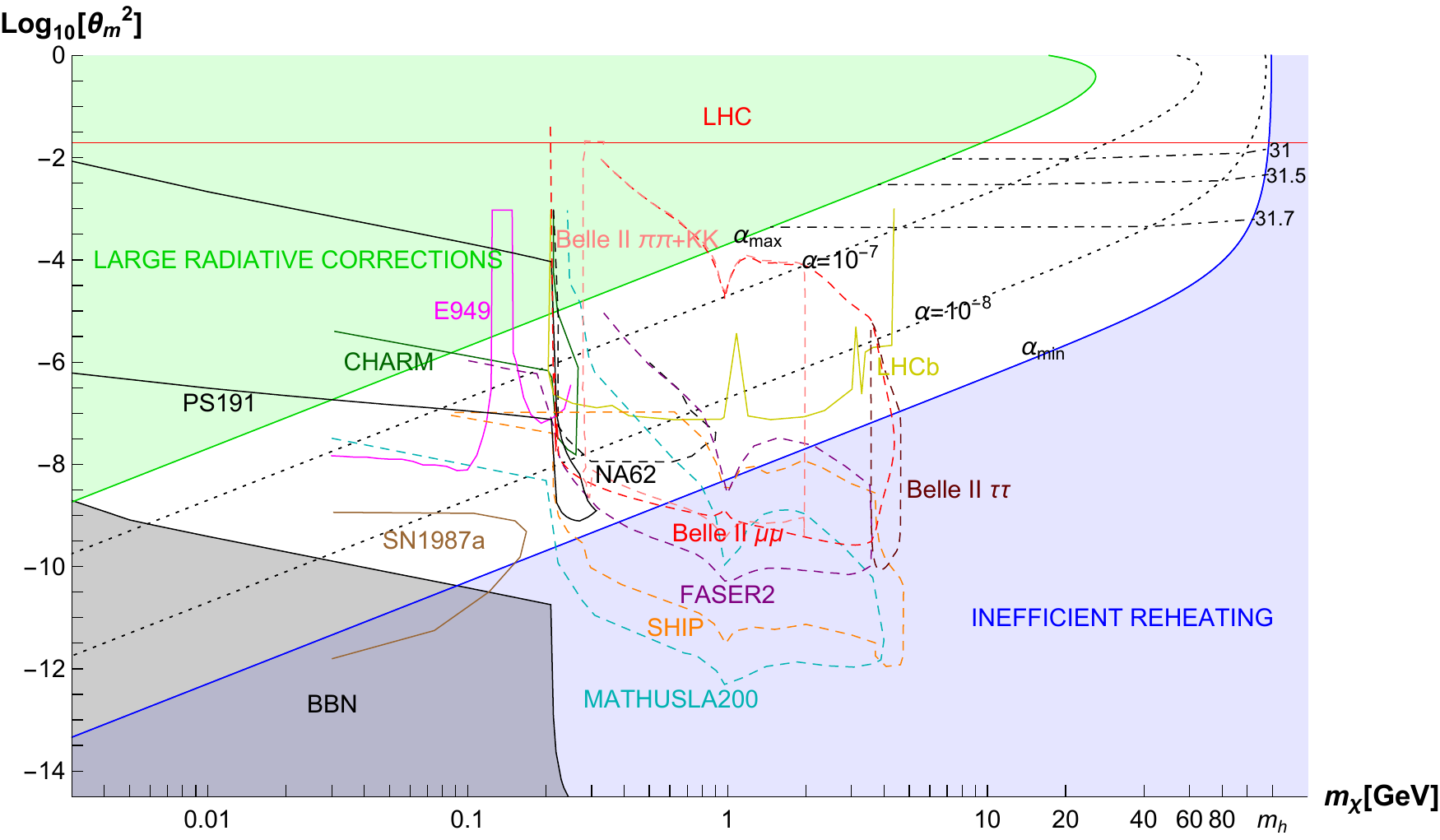}
         \caption{$\beta=10^{-12}$}
         \label{fig:mTb13}
     \end{subfigure}
      \caption{\setstretch{1.0}\small{The plots give the parameter space of the squared mixing angle, $\text{Log}_{10}[\theta_{\text{m}}^{2}]$, against the inflaton mass $m_{\chi}$, for $\beta=10^{-9}$ (top) and $\beta=10^{-12}$ (bottom). The shaded regions are theoretical bounds: (i) Bottom-right (blue): inefficient reheating, $T_{\text{reh}}<T_{EW}$ (Table \ref{tab:ba}); (ii) Bottom-left (grey): inflaton decay after BBN, $\Gamma_{\chi}>t_{\text{BBN}}$ where $t_{\text{BBN}}\sim 1$s (Table \ref{Tab:theta}); (iii) Top-left (green): large radiative corrections to the inflationary potential, $\alpha^{2}>0.1\beta$ (Table \ref{tab:ba}). The full and dashed lines are existing and future experimental bounds. The LHC bound is the $1\sigma$ bound from the ATLAS and CMS Higgs' signal strength measurements, given by $(\ref{eq:theta1})$. All other experimental bounds are taken from \cite{LHCb:2016awg,LHCb:2015nkv,E949:2008btt,Winkler_2019,Turner:1987by,PhysRevD.36.2201,PhysRevD.39.1020,PhysRevD.82.113008,Beacham_2019,Filimonova_2020,Gorbunov_2021}. Dotted lines are curves of constant $\alpha$, and the (almost) horizontal dot-dashed lines are isocurves of the trilinear Higgs coupling, $\lambda v \cos^{3}\theta_{\text{m}}$.}}
     \label{fig:h}
\end{figure}

We obtain the $m_{\chi}-\theta_{\text{m}}^{2}$ parameter space for a given value of the inflaton self-coupling, $\beta$, which is plotted in Figure \ref{fig:h} for $\beta=10^{-9}$ (top) and $\beta=10^{-12}$ (bottom). The shaded regions define theoretical bounds, while solid and dashed lines define current and future experimental bounds respectively. The dotted black curves are lines of constant $\alpha$, and the solid green/blue lines define the boundaries for maximum/minimum $\alpha$, which are given in Table \ref{tab:ba}. The parameter space is fully closed by the solid red line, which gives the 1$\sigma$ bound from ATLAS and CMS measurements of the Higgs signal stength (\ref{eq:theta1}), and the shaded grey region, which eliminates the parameter space where $\Gamma_{\chi}\leq 1$s to ensure inflaton decay prior to BBN (Table \ref{Tab:theta}). The analytical approximation of the maximum inflaton mass (\ref{eq:mxmax}) is in agreement with the results in Figure \ref{fig:h}, which show that the maximum inflaton mass approaches the Higgs boson mass for smallest $\alpha$ and is suppressed for larger $\alpha$. 

Inflatons with $m_{\chi}\sim (1 - m_{h})\text{ GeV}$ have larger mixing angles and so future precision experiments may observe suppressed production cross-sections and tri-linear/quartic Higgs couplings with respect to the SM expectation. The (almost) horizontal dot-dashed lines in Figure \ref{fig:h} are isocurves of the tri-linear Higgs coupling, $\lambda v \cos^{3}\theta_{\text{m}}$, which are suppressed by up to $2\%$ with respect to the SM expected value $(\lambda_{\text{SM}}v=31.7)$. However, fully disentangling the parameter space will require multiple particle physics observables, including the Higgs' couplings and signal strength, as well as a tighter bound on the invisible decay width, corresponding to $\Gamma_{h\rightarrow\chi\chi}$ (\ref{eq:gammahxx}) in our model. Additionally, a measurement of the tensor-to-scalar ratio would allow us to fix $\beta$. 

Because of an increase in motivation to search for new physics in the hidden sector, particles with sub-EW mass that are very weakly-coupled to the SM, there are plans for a large number of future experiments to probe regions beyond the reach of the Large Hadron Collider (LHC) \cite{Beacham_2019}. The testability of our model would greatly benefit from these proposals, as with a combination of experiments, our parameter space could be accessible down to $\theta_{m}^{2}=O(10^{-11})$ for inflaton masses in the range $m_{\chi}\sim (0.03-5)$ GeV, as shown by the dashed coloured contours in Figure \ref{fig:h}.

\section{Discussion and conclusion}
Our model minimally extends the SM via the addition of a single scalar inflaton field with a Z$_{2}$-symmetric potential. By doing so, we are able to incorporate mechanisms for inflation via the quartic scalar coupling, efficient reheating via the scalar-Higgs portal, and symmetry breaking in the Higgs and scalar sectors by the addition of two negative mass terms. 

The addition of the scalar-Higgs portal rotates the mass basis of our model with respect to the gauge basis. As a result, the Higgs boson mass does not define uniquely its quartic coupling and so the stability of the EW vacuum is determined not only by the measurements of SM parameters, but also by the properties of the inflaton field. In order to constrain our model, we instead use the argument that an eternally accelerating Universe is impossible \cite{dvali2014quantum,Dvali_2017} to motivate the need of a metastable EW vacuum. However this is problematic during inflation, as here if $\lambda$ becomes negative for an inflationary attractor that does not align along $\Phi_{b}=0$, it will break down. Only if $\lambda$ is finely tuned close to the stability bound can this be avoided. We therefore proceed to analyse the case where fine-tuning is not necessary: in the parameter space where the inflationary attractor solution is $\Phi_{b}=0$, given by $\alpha>0 \ \& \ 2|\alpha|>\beta$. Additionally, this region of parameter space benefits from generating a large effective mass of the Higgs field during inflation, which exceed the Hubble expansion rate. As a result, the instability scale of the Higgs' potential exceeds the amplitude of the Higgs' inflationary-enhanced quantum fluctuations, thereby preventing the EW vacuum from being destabilized. 

The range of the inflaton's self-coupling, $\beta$, is determined by the CMB's amplitude of scalar perturbations and the upper-bound on the tensor-to-scalar ratio. For a given value of $\beta$, the range of the inflaton-Higgs coupling, $\alpha$, is bounded from below, to ensure reheating temperatures exceeds the EW symmetry breaking scale, and from above, so quantum corrections to the inflationary potential are sufficiently small. The maximum inflaton mass approaches $m_{h}$ for $\alpha^{2}/\beta\leq O(10^{-4})$, and is suppressed for $\alpha^{2}/\beta\geq O(10^{-2})$. We require the inflaton to have a non-zero VEV so that post-reheating they are depleted via decay into SM particles. Inflatons with zero VEV are stable and would overclose the Universe. We require inflatons to decay prior BBN due to the tightly constrained number of additional degrees of freedom at this epoch, thereby constraining inflatons with $m_{\chi}=O(0.001-0.1)$ GeV to have $\theta_{\text{m}}^{2}\geq O(10^{-9}-10^{-11})$. 

To be consistent with the observed Higgs' signal strength results, there is an upper-bound on the inflaton-Higgs mixing angle, $\theta_{\text{m}}\leq0.14$, which translates to a lower bound on the Higgs self-coupling, $\lambda(m_{t})\geq 0.123$. Within this bound, we evade additional cuts to the parameter space from reheating, as for $\lambda(m_{t})>0.120$ the EW vacuum is stable up to $T=O(M_{\text{P}})$. Note that the interesting metastable region in our model $(0.123\leq \lambda\lesssim\lambda_{\text{SM}})$ exists for any value of top quark mass within reasonable experimental errors given in \eqref{eq:mt}. Even if the top quark is light enough for the SM to be stable, this would just reduce the stability bound (\ref{eq:stab}) and only slightly shrink the allowed region in Figures \ref{fig:mTb9} and \ref{fig:mTb13}.

To conclude, the parameter space for $\alpha>0 \ \& \  2|\alpha|>\beta$, which evades all current experimental, cosmological and stability constraints required for a metastable vacuum, spans light inflaton with masses $O(10^{-3}-m_{h}) \text{ GeV}$ and mixing angles $\theta_{m}^{2}=O(10^{-11}-10^{-2})$. We have a rich set of experimental probes from particle physics and cosmology to fully study our model, which requires a multitude of observables: from measurements of the Higgs' couplings, invisible decay width and production cross-section, and a measurement of the tensor-to-scalar ratio. Future upgrades of current experiments may be sensitive enough to observe suppressed linear and trilinear couplings Higgs couplings with respect to the SM in the parameter space of larger mixing angles, $\theta_{m}^{2}\geq O(10^{-4})$, where inflatons have masses $m_{\chi}\gtrsim1\text{ GeV}$. However, accessibility to the parameter space of very weakly-coupled inflatons is dependent on the planned proposals of new experiments \cite{Beacham_2019} that target the hidden sector (namely MATHUSLA200, SHIP, and FASER2). The experimental testability of our model would greatly benefit from a combination of these experiments, which can probe mixing angles down to $\theta_{m}^{2}=O(10^{-11})$ and inflatons with masses $m_{\chi}\sim(0.03-5)$ GeV.

\begin{acknowledgments}\
    The authors of the paper are grateful to S. Westhoff for valuable discussions. The work is supported in part by the Lancaster–Manchester–Sheffield Consortium for Fundamental Physics, under STFC research Grant No. ST/P000800/1 and ST/T001038/1.
\end{acknowledgments}
\bibliography{eprintcontrol,bibfile}

\end{document}